\documentclass[twocolumn,pra,aps,showpacs,superscriptaddress,floatfix,showkeys,nofootinbib,10pt]{revtex4-2}

\usepackage[T1]{fontenc}
\usepackage{amsmath,amsfonts,amssymb,amsthm}
\usepackage{graphicx}
\usepackage{dsfont}
\usepackage[toc,page]{appendix}
\usepackage{subfigure}
\usepackage{bbm}
\usepackage{IEEEtrantools}
\usepackage[inline]{enumitem}
\usepackage{framed}
\usepackage{multirow}

\usepackage{tikz}
\usetikzlibrary{arrows.meta}

\newcommand{\ket}[1]{ | \, #1 \rangle} \newcommand{\bra}[1]{ \langle #1 \, |}

\newcommand{\be}{\begin{equation}} \newcommand{\ee}{\end{equation}}
\newcommand{\ba}{\begin{aligned}} \newcommand{\ea}{\end{aligned}}

\DeclareMathOperator{\Tr}{Tr}

\DeclareRobustCommand\openone{\leavevmode\hbox{\small1\normalsize\kern-.33em1}}%

\newcommand\bigforall{\mbox{\Large $\mathsurround=0pt\forall$}}

\newtheorem{prop}{Proposition}

\newtheorem{obs}{Observation}

\begin{document}

\title{Device-independent Shannon entropy certification}

\author{Robert Okula} \email{rbrt.okula@gmail.com}
\affiliation{Department of Algorithms and System Modeling, Faculty of Electronics, Telecommunications and Informatics, Gda\'{n}sk University of Technology, Poland}
\affiliation{Department of Physics, Stockholm University, S-10691 Stockholm, Sweden} 

\author{Piotr Mironowicz} \email{piotr.mironowicz@gmail.com}
\affiliation{Department of Algorithms and System Modeling, Faculty of Electronics, Telecommunications and Informatics, Gda\'{n}sk University of Technology, Poland}
\affiliation{Center for Theoretical Physics, Polish Academy of Sciences, Aleja Lotników 32/46, 02-668 Warsaw, Poland}

\date{\today}

\begin{abstract}
Quantum technologies promise information processing and communication technology advancements, including random number generation (RNG). Using Bell inequalities, a user of a quantum RNG hardware can certify that the values provided by an untrusted device are truly random. This problem has been extensively studied for von Neumann and min-entropy as a measure of randomness. However, in this paper, we analyze the feasibility of such verification for Shannon entropy. We investigate how the usability of various Bell inequalities differs depending on the presence of noise. Moreover, we present the benefit of certification for Shannon compared to min-entropy, as well as the tight analytical lower bound for Shannon entropy in randomness certification.
\end{abstract}


\maketitle

\section{Introduction} \label{sec:introduction}
The field of computer science and information technology has long been struggling with the problem of random number generation (RNG). The ability to provide truly nondeterministic numbers is crucial for the security of most modern cryptography systems \cite{schoen-rsa, cryptoeprint:2012/064, 10.1007/978-3-642-40576-1_7}, as well as physical or biological simulations \cite{jcc21638}, machine learning \cite{KOIVU2022117938, ALTARABICHI2024120500}, among others. Currently, the most widely utilized method is based purely on the predictable manipulation of an integer value of the seed, which makes it deterministic and therefore only provides an illusion of randomness (we call it a pseudo-random number generator). Although there are numerous pseudo-RNG solutions available, which differ in the quality of generated data \cite{Vattulainen_1995, ullah2019efficientsecuresubstitutionbox, Falcioni_2005}, they still do not provide real random numbers or a random stream of bits.

Some solutions use processes that represent certain chaotic properties (such as the atmospheric noise \cite{randomorg} or radioactive decay \cite{hotbits}). However, even if these phenomena might present a high level of entropy in most situations, they are not provably random, which creates a possibility of even partial prediction of the random value, reducing its final entropy. Furthermore, such sources demand the usage of certain applications or websites \cite{randomorg, hotbits, nistbeacons}, which process these phenomena, often without any reliable proof that the claimed source of randomness has been used or that it has not been manipulated. The users have to trust these suites or build their own advanced infrastructure.

The diversity of RNG or pseudo-RNG methods demands a mechanism for their verification and comparison. The National Institute of Standards and Technology (NIST) developed and summarized a set of statistical tests, which allow for characterization and selection of appropriate (pseudo-)RNG solutions \cite{10.5555/2206233}. However, the authors also pointed out that these tests are just as useful as the first step if the generator is suitable for a certain usage; they do not replace proper cryptoanalysis or other methods of randomness certification.

In 1964, John Bell \cite{PhysicsPhysiqueFizika.1.195} demonstrated that quantum mechanics is incompatible with local hidden-variable theories proposed, among others, in the context of the Einstein-Podolsky-Rosen paradox \cite{PhysRev.47.777}. The correlations predicted for entangled particles were shown to exceed the bounds allowed for any local realistic model, revealing the intrinsic indeterminacy of quantum measurement outcomes.

The more non-deterministic the sources that are available to us when generating randomness, the better. Thus, the probabilistic properties of quantum phenomena constitute a viable tool for randomness generation, due to their intrinsic unpredictability. To this day, many quantum-based RNG (QRNG) have been proposed \cite{RevModPhys.89.015004}, including, but not limiting to, entanglement-based \cite{xavier2010practicalrandomnumbergeneration, Kwon_2009, Xu:16, MONGIA2024129954, Seguinard_2023}, based on photon-number statistics \cite{Furst:10, PhysRevA.83.023820} or on binary measurement of coherent states \cite{Wei:09, PhysRevApplied.22.044041}. The existence of QRNG is crucial for the security of other means of quantum cryptography \cite{RevModPhys.74.145, Pirandola_2020}, such as quantum key distribution \cite{Bennett_2014, PhysRevLett.67.661, 8527822} or quantum digital signatures \cite{gottesman2001quantumdigitalsignatures}.

In practice, even if we can be sure that a randomness protocol, quantum or not, produced some amount of randomness, in most cases the source does not give us a uniform probability distribution, but some result that might even have a very high entropy, but is still biased. Therefore, certain methods called \textit{extractors} are used, which transform a stream of biased numbers into shorter streams that are almost uniformly distributed \cite{Sunar2009}.

Given that, an interesting problem arises -- QRNG validation and the maximization of entropy. However, the sheer nature of random numbers prohibits the verification methods that do not involve probability statistics or the study of parameters of quantum phenomena. The ideal scenario for users of such a device would be the achievement of unconditional, device-independent (DI) \cite{Ac_n_2007} security, whereby the randomness can be fully certified solely from the observed input-output statistics of the device, independent of its internal implementation.

To tackle this problem, in this paper, we propose a DI QRNG certification method that uses Shannon entropy as the randomness measure, and we show which Bell inequalities certify the highest amount of randomness for different levels of noise.

The paper is organized as follows: In the latter part of Section~\ref{sec:introduction}, we discuss the idea for quantum randomness certification, as well as a semi-definite programming (SDP) technique that we use to obtain a probability bound, called the Navascués-Pironio-Acín (NPA) hierarchy. In Section~\ref{sec:exp}, we introduce a method that we use to describe Bell inequalities that can be used for certification, and in Section~\ref{sec:method}, we present our DI QRNG certification method that uses Shannon entropy, including the developed numerical protocol. The numerical certification results for the previously known Bell inequalities are presented in Section~\ref{sec:results}. Section~\ref{sec:discussion} contains reflections on the course of Shannon entropy plots and the results for the certification with additional constraints, as well as entropy certification for three-dimensional quantum states, based on the BM functional. Furthermore, we establish a tight analytical lower bound for the value of Shannon entropy. We also present the best randomness certification method for the noise level ranges.

\subsection{Shannon entropy} \label{ssec:shannon}
Information entropy quantifies the level of uncertainty of the outcome of a given operation. It was initially introduced in \cite{shannon}, where the mathematical quantification of information was established. Through the formulation of this measure, the core principles of information theory were established, providing a foundational framework for, among others, reliable communication over noisy channels. It has been defined as follows: for a given discrete probability distribution $P = \left( p_1, p_2, \ldots, p_n \right)$,
\begin{equation}
    \label{eq:shannon}
    H(P) \equiv - \sum_{i=1}^{n} p_i \log_2 p_i.
\end{equation}
This type of entropy describes the average unpredictability of a certain outcome, but it can also be used to measure a lower bound on the average number of bits needed to represent the information. Another type of information entropy, which corresponds to the unpredictability of the most probable event or result, is defined as:
\begin{equation}
    \label{eq:minentropy}
    H_{min}(P) \equiv - \log_2 \left( \max(p_i) \right).
\end{equation}
A lower bound on Shannon entropy of a given random variable is provided by min-entropy, which was investigated in the context of randomness certification in \cite{pironio_random_2010, Mironowicz_2013, Um2013, Berta_2014, PhysRevA.90.052327, Passaro_2015, Seguinard_2023}. It is caused by the fact that the Shannon entropy yields information about the distribution, taking the average over many independent samples, but min-entropy takes only one, worst-case option, directly bounding the probability that an adversary can correctly guess the value of the random variable.

If the samples are produced independently, a property called the Asymptotic Equipartition Property (AEP) holds \cite{10.1007/978-3-319-31875-2_7}. The min-entropy per bit in a sequence of independent and identically distributed random variables $X_1, X_2, \ldots, X_n$ converges to the Shannon entropy $H$ (for $n \rightarrow \infty$) of the sequence:
\begin{equation}\label{aep}
	-\frac{1}{n} \log P_{X_1, \ldots, X_n} \rightarrow H(X).
\end{equation}
In essence, this property leads to the conclusion that the more samples we have (for instance, the larger set of QRNG-based values we analyze), the more relevant a proper certification of Shannon entropy becomes. Indeed, for a generation of a single sample, we need to assume that the worst possible scenario can happen, and therefore, min-entropy must be considered the certified lower bound. As the number of samples grows, the likelihood that most individual sample values cluster near their respective lower bounds decreases. This implies that, for sufficiently large sample sizes, the aggregate entropy of the set will tend to deviate from the min-entropy-based lower bound. Min-entropy certification, therefore, becomes increasingly conservative, which realistically underestimates the amount of randomness obtained from the system. For a large set of samples, the entropy of the whole set tends to converge to the average value, which is the Shannon entropy.

The problem of reliable Shannon entropy estimation for QRNG processes has been initially studied in \cite{vanhimbeeck2019correlationsrandomnessgenerationbased}. They have shown a memoryless semi-DI QRNG scheme, as well as a generic method for semi-DI QRNG rate computation. They have, however, made some assumptions that reduce the applicability of their method, e.g., they conduct a linearization of the entropy function and, based on that linearization, they optimize the linear constraints over the set of quantum behaviors that describe the correlations.

Our method, although also uses Shannon entropy for randomness estimation, does not require meeting these limitations -- it can be used for full DI protocols, it does not require a linearization (it relies on non-linear optimization, bounded by the SDP-obtained lower and upper bounds, instead), and it uses a set of constraints that is based purely on the probability distribution of the correlations.

\begin{figure}[htbp]
	\includegraphics[width=\linewidth]{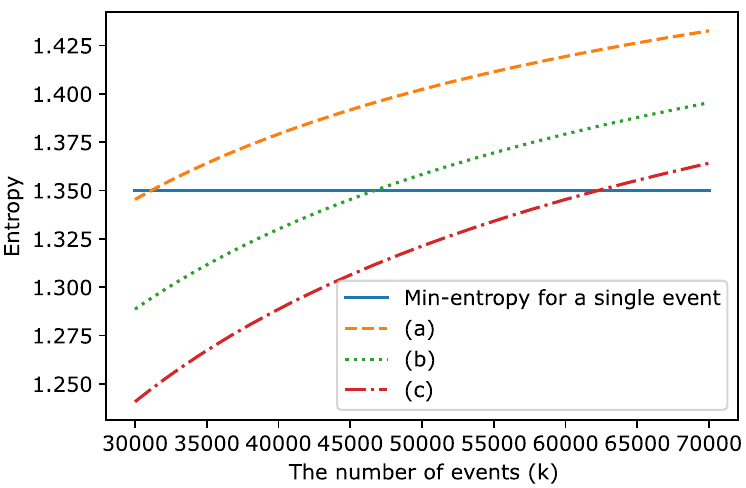}
	\caption{(color online) Shannon entropy-based min-entropy sequence lower bounds based on \eqref{eq:prep1}, per event in the sequence, compared to the min-entropy for a single event. Three statistical distance bounds $\varepsilon$ has been analyzed: (a) $\varepsilon = 10^{-8}$; (b) $\varepsilon = 10^{-12}$; (c) $\varepsilon = 10^{-16}$. In (a), for $k \approx 30k$ (and $k \approx 45k$ in (b), and $k \approx 62.5k$ for (c)), this lower bound (divided per event) exceeds the maximal single-event min-entropy value for CHSH.}
	\label{fig:hol}
\end{figure}

Estimated Shannon entropy can also be used to determine the effective number of bits that can be extracted from the quantum source. Randomness extractors use the min-entropy to quantify how much true unpredictability a weak random source has, so they can guarantee that their output is close to uniform only when the source has at least that much entropy. In \cite{10.1007/11681878_23} it has been shown, however, that for several independent samples $k$, the conditional min-entropy of the source $X^k$, given the setting $Y^k$, is close to the Shannon entropy. For any statistical distance $\varepsilon$ there exists a distribution $P({\overline{X}|\overline{Y}})$, which is at most $\varepsilon/2$ from $(X^k, Y^k)$ and satisfies:
\begin{equation}
\label{eq:prep1}
	H_{min}(\overline{X}|\overline{Y}) \ge kH(X|Y) - 6 \sqrt{k \log_2 \frac{1}{\varepsilon} \log_2 |X|}.
\end{equation}
This establishes a quantitative connection between Shannon entropy and min-entropy: for sufficiently long sequences of independent samples, the conditional min-entropy of the joint distribution grows approximately linearly with the conditional Shannon entropy. Consequently, the amount of extractable randomness from the entire sequence is governed more by the average uncertainty (as captured by Shannon entropy) than by the min-entropy of any individual event. This establishes a connection between Shannon entropy and min-entropy and shows that for an appropriately long sequence of events, the extractable min-entropy for the whole string is more related to the Shannon entropy (corresponding to the mean randomness value) than to the min-entropy for a single event. We present this phenomenon in Figure \ref{fig:hol}.

Furthermore, it has also been shown in \cite{10.1007/11681878_23}, that when the bound \eqref{eq:prep1} guarantees sufficient conditional min-entropy, a standard extraction method can be applied, e.g. the Leftover Hash Lemma, where Shannon entropy determines how many bits can be extracted, as the extractor's output lengths is proportional to $m = kH(X|Y) - 6 \sqrt{k \log_2 \frac{1}{\varepsilon} \log_2 |X|}$ and thus to the value of Shannon entropy. Similar analysis has also been presented in \cite{10.1007/978-3-319-31875-2_7}, where the upper bound on bits extractable from Shannon sources has been obtained:
\begin{equation}
	N  = H k - \Theta \left( \sqrt{\log_2 (1/\varepsilon) hk} \right),
\end{equation}
The constant under $\Theta \left( \cdot \right)$ depends on the source, similarily to $\log_2 |X|$ in \cite{10.1007/11681878_23}. The key assumption, however, is that the source is memoryless, an assumption that also applies to our setting.

\subsection{Navascués-Pironio-Acín hierarchy} \label{ssec:npa}
Introduced in \cite{PhysRevLett.98.010401, Navascués_2008}, the NPA method consists of an infinite hierarchy of conditions satisfied by any set of quantum correlations. Then, every level in the hierarchy is converted into an SDP problem, which is then solvable numerically. The method can be used in the context of quantum mechanics as follows. We operate on probability distributions for certain projective measurement operators, with settings $x$ and $y$ and results $a$ and $b$ for Alice ($M^{x}_a$) and Bob ($M^{y}_b$) respectively, such that $P(a,b|x,y) = Tr(\rho(M^{x}_a \otimes M^{y}_b))$, where $\rho$ is a quantum state, shared between the parties. In our case, the measurements are binary, thus $a, b \in \{-1, 1\}$. If the parties used quantum states other than qubits (e.g., qutrits), the number of possible outcomes would be higher.
This method allows for effective numerical study of quantum correlations, including quantum randomness. It is especially valuable for analyzing complex quantum systems, offering key insights into quantum correlations and phenomena. It has been successfully used to find min-entropy lower bounds \cite{Mironowicz_2013}, as well as for von Neumann entropy \cite{koßmann2024boundingconditionalvonneumannentropy}. As this method is a form of relaxation, it only provides bounds on the exact solution, which can then be utilized to constrain other calculations.

\section{Device-independent randomness extraction} \label{sec:exp}
Let us consider the case with two independent and spatially separated parties, Alice and Bob, and a third party that functions as a referee. The referee shares an entangled state with both parties (e.g., a state in the form of two entangled photons, one for each party). Alice and Bob then independently measure their particles. For example, they can use two sets of binary measurements, $\{M^{1}_a, M^{2}_a\}$ and $\{M^{1}_b, M^{2}_b\}$. If the measurement outcome on Alice's particle cannot be influenced by actions or settings performed on another and spatially separated Bob's particle, an inequality:
\begin{equation}
    \label{eq:oldchsh}
    | \langle M^{1}_a M^{1}_b \rangle + \langle M^{1}_a M^{2}_b \rangle + \langle M^{2}_a M^{1}_b \rangle - \langle M^{2}_a M^{2}_b \rangle | \leq 2,
\end{equation}
($\langle M^{x}_a M^{y}_b \rangle$ denotes the expectation value between Alice’s and Bob’s measurement outcomes when they use settings $x$ and $y$) will always be true. However, if the particles are in any way codependent, the value of the expression will be higher than 2. The inequality in the form of \eqref{eq:oldchsh} is called a Bell inequality, and the expression in the inequality is called the Bell expression. The one presented in \eqref{eq:oldchsh} is called Clauser-Horne-Shimony-Holt (CHSH) \cite{PhysRevLett.23.880} and is probably the most well-known and proliferated Bell inequality. Its classical (or local) bound equals 2, but its quantum bound, the value that this expression yields for the maximal entanglement, equals $2 \sqrt{2}$ \cite{cirelson_quantum_1980}. We call the quantum bound Tsirelson's bound (we mark it as $B_{max}$), which has an individual value for each Bell expression.

To simplify the notation of the Bell expression, we define the correlation operator $C(x,y)$ as:
\begin{equation}
	C(x,y) = 4 \Pi_+^{0, x} \Pi_+^{1, y} - 2 \Pi_+^{0, x} - 2 \Pi_+^{1, y} + \mathbbm{1},
\end{equation}
where $\Pi_{+}^{0, x}$ ($\Pi_{-}^{0, x}$) is Alice's projector on results $+1$ ($-1$) and analogously for Bob $\Pi_{+}^{1, y}$ ($\Pi_{-}^{1, y}$), as in our main analysis, the results of each separated measurements are binary (although in general, they do not have to be, as presented in Section \ref{ssec:3dscenario}). Given that, for a state $\rho$ of the entangled pair:
\begin{equation}
\begin{split}
	\Tr(\rho C(x,y)) = P(a = -1, b = -1|x,y)\\
	- P(a = -1, b = 1|x,y)\\
	- P(a = 1, b = -1|x,y)\\
	+ P(a = 1, b = 1|x,y),
\end{split}
\end{equation}
which constitutes the correlations between the binary results obtained by Alice with Bob's results, for the settings pair $(x,y)$.

The party that is willing to verify the unpredictable nature of the generated randomness can use the violation of any Bell inequality as a certificate of randomness \cite{colbeck2011quantumrelativisticprotocolssecure}. This method, introduced by \cite{pironio_random_2010}, relies on the assumption that two correlated systems are separated. It also requires an initial random seed to perform statistical tests verifying the violation of the Bell inequality. Therefore, it is best classified as a randomness expansion protocol \cite{colbeck2011quantumrelativisticprotocolssecure}, with randomness quantified using the min-entropy measure. It was also shown that this method of randomness certification is, in principle, device-independent \cite{Pironio_2013}. This idea has been further investigated experimentally in e.g. \cite{Wang_2023, Li_2021, zhang2024, piveteau2024}.

The randomness certification method presented in \cite{pironio_random_2010} is based on the CHSH inequality. However, since this is not the only available option, the user or vendor of the QRNG device may employ alternative Bell inequalities. In this case, a question arises -- which one should they use to maximize the amount of randomness, or in other words, which one is going to maximize the unpredictability of the random bit obtained?

In our analysis, we are constrained to the bare input and output of Alice's and Bob's devices: we do not know how the devices work internally, we do not know how the measurements are conducted (if they are done correctly or if the results are biased in any way), as we want to work in the DI scenario. We can, however, make some assumptions about the externals of the setup. We can assume that both devices are spatially separated from each other (which means that there is no way for any signal transmission between the two parties that might convey information about the state of one, which could be used to manipulate the results to make them statistically imitate entanglement). Furthermore, we should also assume that the setup does not communicate with the outside world in any other way than by controlled input and output of the QRNG protocol.

For a two-qubit system, the largest number of random bits obtained will be 2; thus, the maximal amount of randomness that can be certified by any protocol is $H = 2$, and it emerges for a uniform probability distribution ($P(a,b|x,y) = 0.25$) for all four possible measurement outcomes.

The objective of the analysis is to obtain the maximal randomness, quantified, in this case, by the maximum value of the Shannon entropy. To achieve that, we have conducted a chain of numerical calculations, at first examining the behavior for a scenario in which both communicating parties measure entangled states.

Let us consider an experiment on a quantum entangled state of two particles, shared previously between Alice and Bob, which are the Werner states:
\begin{equation}
\rho = p \frac{\mathbbm{1}}{4} + (1 - p) \ket{\psi^+} \bra{\psi^+},
\end{equation}
where $\bra{\psi^+}$ is a maximally entangled state and $p$ is the noise level. The parties separately choose observables (settings) and perform measurements on their particles. Only a handful of setting pairs will, however, maximize the value of Shannon entropy for given Bell inequalities.

To guarantee a valid DI randomness expansion, a protocol called spot-checking should be used \cite{10.1007/978-3-642-40328-6_33,Shalm2021}. It is used to verify the integrity of quantum devices without having to test every round, ensuring that the observed correlations genuinely reflect quantum behavior. It provides strong security guarantees because the random and unpredictable selection of test rounds prevents adversarial devices from knowing when they are being checked, making cheating statistically detectable. For a Bell inequality $G$, a trusted third party determines randomly (randomly chooses $t_i=\{ 0, 1 \}$), with a small probability $0 < q < 1$ (the longer the output we want to obtain, the smaller $q$ needs to be, e.g. in \cite{doi:10.1137/15M1044333} $q = \frac{\log^2 N}{N}$ has been used), whether a trial will be one of two options: testing the value of Bell expression or randomness generation.
%\begin{enumerate}
%\item if $t_i = 0$, Alice and Bob use a fixed input $(x^{t_i=0}, y^{t_i=0})$ and obtain an expected Bell violation value $B^{t_i = 0}$;
%\item if $t_i = 1$, Alice and Bob conduct a Bell experiment and record the Bell violation value $B^{t_i = 1}_i$.
%\end{enumerate}
%For the output length $N$ then, if the condition \cite{doi:10.1137/15M1044333}:
%\begin{equation}
%\sum_{i = 1, \ldots, N; t_i = 1} B^{t_i=1}_i \ge q N B^{t_i=0},
%\end{equation}
%is met, the procedure succeeds.

\section{Methods for Shannon entropy certification}
\label{sec:method}
The standard (e.g. \cite{colbeck2011quantumrelativisticprotocolssecure, pironio_random_2010, Colbeck_2011, Mironowicz_2013}) approach to the Bell certification is based on independent optimization of the outcomes: for a given Bell inequality, we determine its value and then maximize the probabilities of individual outcomes. However, when every outcome is maximized independently, we omit the natural constraints that result from the probability properties -- all outcomes cannot be maximized at the same time, and maximization of one outcome will affect the value of the remaining.

This fact does not interfere with the certification on the value of min-entropy, as in that situation, we only utilize the highest probability. However, for the certification on the value of Shannon entropy, even if a single probability is high, the result is influenced by the total probability distribution. The quantum behavior that maximizes min-entropy might likely not be the one that maximizes the value of Shannon entropy.

Thus, a need for a different method arises. A solution based on NPA can be used to establish the bounds on probabilities allowed by quantum mechanics. These bounds can be used as constraints on the probability distribution. Essentially, we can use NPA to lower and upper bound any linear probability equation, and these results can be used as constraints on the non-linear optimization of the Shannon entropy.

In other words, for any Bell inequality, we can establish a set of linear expressions $\{ L_i \}_i$:
\begin{equation}
\begin{aligned}
	L_i(P, x_0, y_0) = \sum_{a,b} c^{(abx_0y_0)}_i P(a,b|x_0,y_0),
\end{aligned}
\end{equation}
where $c^{(abx_0y_0)}_i$ is a constant and $(x_0,y_0)$ are the spot-settings, which are specified for each optimization. For every linear expression, we can now conduct an NPA optimization. Consider the following optimization problem:
\begin{equation}
\label{eq:npalinear}
\begin{aligned}
\max_P \quad & L_i(P, x_0, y_0)\\
	\textrm{subject to} \quad & G(P) - (1-p) B_{max} = 0, \\
\end{aligned}
\end{equation}
where $G(P)$ is the value of the Bell expression for a given probability distribution and $B_{max}$ is the inequality's Tsirelson's bound. Using the NPA method, we solve \eqref{eq:npalinear}, receiving the expression's upper bound (to obtain a lower bound, we need to minimize). In our case, the optimization has been computed using a $Q_2$ level of the NPA hierarchy.

To be precise, we conduct the NPA-based optimization for the set of linear expressions \eqref{eq:npalinear} for the constant coefficients $c^{(abxy)}_i$ that equal:
\begin{equation}
c^{(-1,-1,x_0, y_0)}_1, c^{(-1,1,x_0, y_0)}_2, c^{(1,-1,x_0, y_0)}_3, c^{(1,1,x_0, y_0)}_4 = 1,
\end{equation}
and the remaining constants $c^{(abxy)}_i = 0$. We progress with the SDP optimization for all possible measurement outcomes ($a = -1$, $b = -1$; $a = 1$, $b = -1$, etc.) and the settings are chosen to maximize entropy with the absence of noise ($p \approx 0$ relates to $H \approx 2$). These cases can be recognized by the lower and upper bounds for each outcome, which are approximately $0.25$.

In the next step, we conduct a non-linear optimization, constrained using the NPA-obtained bounds:
\begin{equation}
\label{eq:nonlinear}
\begin{aligned}
\min_{P} \quad & - \sum_{i \le N_P} S(P_i) \\
\textrm{subject to} \quad & \bigforall_{i \le N_P}, l_i \le P_i \le u_i,
\end{aligned}
\end{equation}
where $l_i$ are the lower bounds, $u_i$ are the upper bounds, $S(P_i)$ is the non-convex optimized function and $N_P$ is the number of the optimized probabilities $P_i$ for the function $S$. In principle, this method can be used for any non-convex problem that can be bounded using the NPA method (thus, the variables are probabilities), which is any problem that requires the characterisation of the correlations that arise from locally measuring a single part of a joint quantum system. In our certification method, the optimized function $S(P_i)$ is the Shannon entropy for a single probability (from the distribution), whereas $P$ is the quantum behavior that minimizes the entropy.

The values of Shannon entropy were obtained using non-linear optimization with the scheme presented in Figure~2. We optimized over three variable probabilities (which we denote as $P_{a,b} = P(a,b|x_0,y_0)$ to simplify the notation): $P_{-1,-1}, P_{-1,1}, P_{1,-1}$ ($P_{1,1} = 1 - P_{-1,-1} - P_{-1,1} - P_{1,-1}$), which correspond to the probability of obtaining each of the allowed outcomes. If the measurement results are not binary, the number of variables equals $n^2 - 1$, where $n$ is the number of outcomes allowed by a local measurement. We have calculated the bounds for every single one of these variables by solving the \eqref{eq:npalinear} optimization problem for the set of linear expressions formed for these single probabilities.

We utilize the COBYQA \cite{rago_thesis, cobyqa} solver that utilizes a trust-region Sequential Quadratic Programming (SQP), which is an iterative optimization method that solves a series of quadratic subproblems to approximate and minimize a nonlinear objective function subject to constraints. A key feature of this method is that it always respects bound constraints throughout the optimization process, which makes it perfect for a problem bounded with NPA relaxation. However, as this problem is non-convex, it has many local minima, where the optimizer may get stuck. To avoid that, we used a global optimization technique called basin-hopping \cite{Wales_1997}, which performs a random perturbation of coordinates with every repetition and accepts the new coordinates based on the minimized function value. In this optimization, we have settled (with trial and error) for 1500 iterations each.

\begin{figure}
\label{fig:alg1}
\begin{framed}
	\textit{Variables}
\begin{description}[font=\sffamily, leftmargin=1cm]
  	\item[P] Probability of measurement outcome.
\end{description}
	\textit{Inputs}
\begin{description}[font=\sffamily, leftmargin=1cm]
  	\item[G] Bell inequality used as a certificate.
  	\item[$B$] Considered Bell expression value.
  	\item[L] Set of linear expressions on the probabilities of results in a form of $L_i(P, x_0, y_0) = \sum_{a,b} c^{(abx_0y_0)}_i P(a,b|x_0,y_0)$.
  	\item[$x_0$,$y_0$] Spot-setting for measurements.
\end{description}
	\vspace{1em}
	\textbf{Protocol}
	\begin{enumerate}[leftmargin=0.4cm]
		\item Using NPA, establish an upper ($u_i$) and lower ($l_i$) bounds of $L_i(P, x_0, y_0)$, subject to $G(P) - B = 0$, where $B = (1-p)B_{max}$ for a noise level $p$.
		\item Using a non-linear optimization, minimize $\sum_{a,b} H(P(a,b|x_0,y_0))$, where every linear expression in a form of \eqref{eq:npalinear} is constrained by $l_i$ and $u_i$ only and $H$ is the objective function (in our case -- Shannon entropy). 
	\end{enumerate}
\end{framed}
\caption{Brief description of our randomness certification method using Shannon entropy.}
\end{figure}

\section{Results for defined inequalities}\label{sec:results}
Using the protocol presented in Section~\ref{sec:method}, we conducted the classification of four Bell expressions for Shannon entropy-based QRNG certification. We were able to evaluate the maximal amount of randomness certifiable, as well as the course of the entropy function, depending on the noise level $p$.

\subsection{Clauser-Horne-Shimony-Holt inequality} \label{ssec:chsh}
Introduced in \cite{PhysRevLett.23.880}, this is the first and therefore the most widely-used example of the Bell inequality. The regular CHSH, however, does not certify two bits of randomness, even for the lack of noise ($p = 0$). It is, however, efficient for $p \ge 0.1$.

By adding one correlation function $C(2,3)$, we can obtain a different operator, which we denote as modified CHSH, or ModCHSH, introduced in \cite{Mironowicz_2013}:
\begin{equation}
\label{eq:modchsh}
    ModCHSH = C(1,2) + C(1,3) + C(2,1) + C(2,2) - C(2,3).
\end{equation}
The Tsirelson's bound \cite{cirelson_quantum_1980} for this Bell inequality is $B_{max}^{ModCHSH} = 1 + 2\sqrt{2}$. We use this protocol for the pair of settings $(x=1,y=1)$.

\begin{figure}[htbp]
	\includegraphics[width=\linewidth]{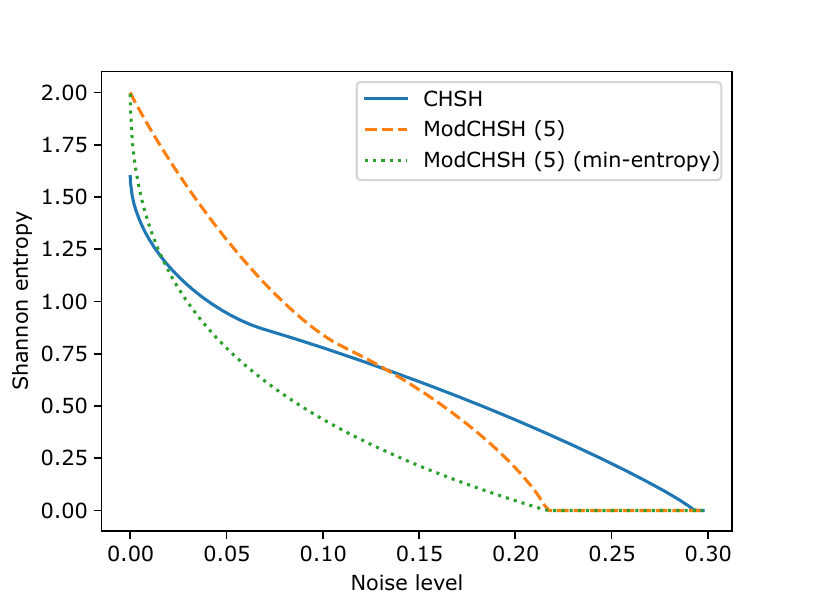}
	\caption{(color online) Lower bounds on Shannon entropy for randomness certification provided by CHSH inequality, as well as \eqref{eq:modchsh}, as a function of the level of noise.}
	\label{fig:ModCHSHCHSH}
\end{figure}

As shown in Figure~\ref{fig:ModCHSHCHSH}, modified CHSH in the form of \eqref{eq:modchsh} indeed leads to the certification of two bits of randomness. However, it does not provide an advantage over classical CHSH for the noise $p \ge 0.1325$. Above this level, the classical CHSH provides higher entropy. In other words, for higher noise, CHSH can certify a higher level of randomness, compared to \eqref{eq:modchsh}.

\subsection{Braunstein-Caves inequalities}
\label{ssec:bcn}
This type of Bell inequality was introduced in \cite{PhysRevLett.61.662}. It is a family of expressions, parametrized by several binary measurement settings for two parties $n \geq 2$:
\begin{equation}
\label{eq:bcn}
\begin{split}
    BC_n = C(1,1) + C(1,2) + C(2,2) + \ldots +\\
    C(n-1,n-1) + C(n-1,n) + C(n,n) - C(n,1)
\end{split}
\end{equation}
The Tsirelson's bound equals $B_{max}^{BC_n} = 2n \cos(\frac{\pi}{2n})$. We have run the optimization for three inequalities that are a part of this family: $BC_3$, $BC_5$, and $BC_7$. We have tried all twelve settings for each inequality, and we have listed those for which two bits of randomness can be certified. For $BC_3$ these are $(x = 1, y = 3)$, for $BC_5$ these are $(x = 4, y = 2)$ or $(x = 1, y = 4)$ and for $BC_7$ these are $(x = 4, y = 1)$ or $(x = 1, y = 5)$. 

\begin{figure}[htbp] \label{fig:BCn}
	\includegraphics[width=\linewidth]{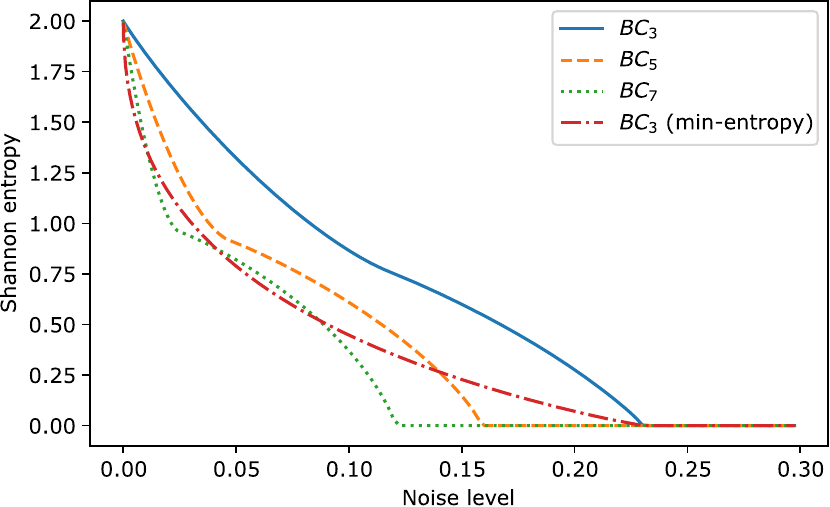}
	\caption{(color online) Lower bounds on Shannon entropy for randomness certification provided by the family of Braunstein-Caves inequalities, as a function of the level of noise.}
\end{figure}

The results for this family of protocols are presented in Figure~\ref{fig:BCn}. They show that the shortest Bell expression, BC3, provides the highest Shannon entropy, followed by BC5 and BC7. However, due to the unusual change of the course of the function, for some noise levels ($p \approx 0.04$), the difference between the BC5 and BC7 is relatively small.

\subsection{$I_1$ and $I_2$} \label{ssec:i12}
In \cite{Mironowicz_2013}, two Bell expressions have been proposed:
\begin{equation}
\label{eq:I1}
\begin{split}
    I_1 = C(1,2) - C(1,3) - C(2,1) - C(2,2)\\
		+ C(3,1) + C(3,3) + C(4,1);
\end{split}
\end{equation}
and
\begin{equation}
\label{eq:I2}
\begin{split}
    I_2 = -C(1,2) + C(1,3) + C(2,1) + C(2,2) + C(2,3)\\
    + C(3,2) - C(3,3) + C(4,1) + C(4,2) + C(4,3).
\end{split}
\end{equation}
The Tsirelson's bound for these are $B_{max}^{I_1} = 1 + 6 \cos(\pi/2)$ and $B_{max}^{I_2} = 2 + 4\sqrt{2}$ respectively. They represent a wider group of inequalities with 4 possible measurement settings for Alice and 3 for Bob. Their min-entropy certification has been thoroughly described in \cite{Mironowicz_2013}, but for the certification of the Shannon entropy, their usefulness has not yet been shown. We evaluate these inequalities for $(x=1,y=1)$ measurement settings, as these lead to the situation where two bits of randomness can be obtained.

\begin{figure}[htbp]
	\includegraphics[width=\linewidth]{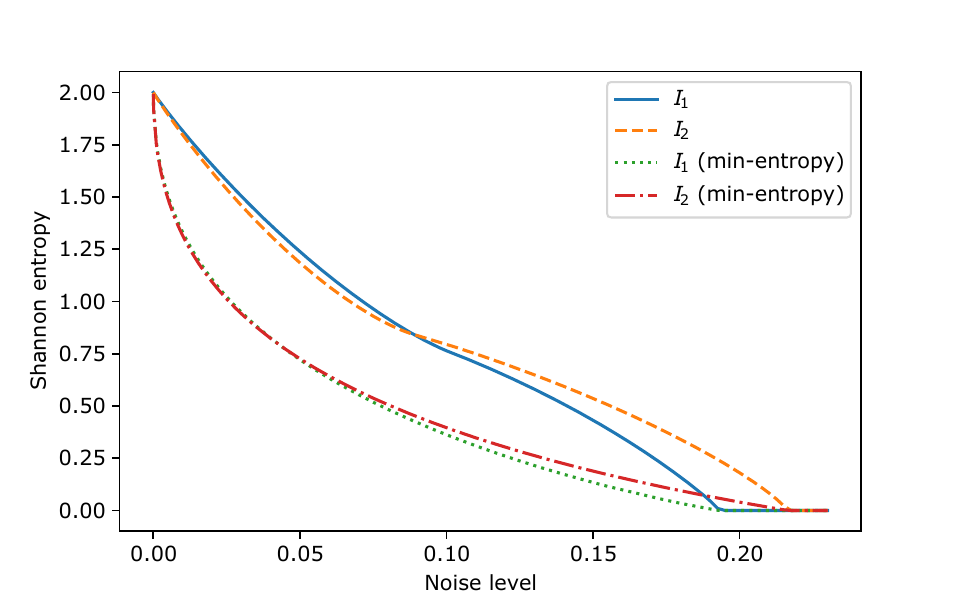}
	\caption{(color online) Lower bounds of Shannon entropy for randomness certification provided by the $I_1$ and $I_2$ inequalities, as a function of the level of noise.}
	\label{fig:I12}
\end{figure}

Both plots presented in Figure~\ref{fig:I12} show that $I_1$ has a small, yet noticeable, advantage over $I_2$ for a smaller amount of noise, whereas this changes at the point where the convexity of the entropy changes. Thus, for larger noise, $I_2$ has a visible (and more influential) advantage; it certifies more randomness.

\subsection{Wooltorton-Brown-Colbeck inequalities} \label{ssec:IJ}
\label{sec:wbc}
The last two families of inequalities investigated in this paper were introduced in \cite{PhysRevLett.129.150403}. The first family, $I_{\delta}$, is defined as:
\begin{equation}
\label{eq:id}
\begin{split}
	I_{\delta} = C(0,0) + \frac{C(0,1) + C(1,0)}{\sin \delta} - \frac{C(1,1)}{\cos 2\delta}
\end{split}
\end{equation}
where $0 \le \delta \le \pi/6$ and Tsirelson's bound equals $B_{max}^{I_{\delta}} = \frac{2 cos^3 \delta}{\cos 2 \delta \frac{1}{\sin \delta}}$. The second one, $J_{\gamma}$, is defined as:
\begin{equation}
\label{eq:jg}
\begin{split}
	J_{\gamma} &= C(0,0) - C(1,1)\\
	&+ (4 \cos^2 \frac{\gamma + \pi}{6} - 1) (C(0,1) + C(1,0)),
\end{split}
\end{equation}
where $0 \le \gamma \le \pi/12$ and the quantum bound equals $B_{max}^{J_{\gamma}} = 8 \cos^3 \frac{\gamma + \pi}{6}$.

\begin{figure}[htbp]
	\includegraphics[width=\linewidth]{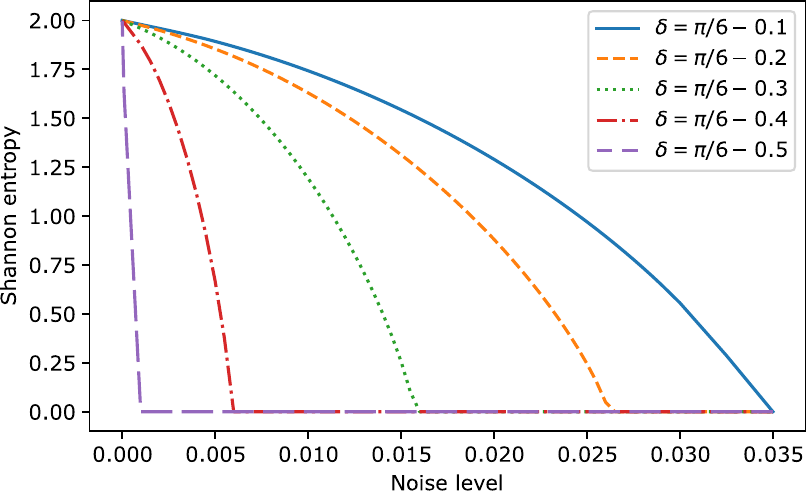}
	\caption{(color online) Lower bounds of Shannon entropy for randomness certification provided by the $I_{\delta}$ (with decreasing values of $\delta$) inequalities, as a function of the level of noise.}
	\label{fig:IDcomparison}
\end{figure}

\begin{table}[htbp]
\begin{tabular}{|l|ll|ll|}
\hline
\multicolumn{1}{|c|}{\multirow{2}{*}{noise level $p$}} & \multicolumn{2}{c|}{$I_{\delta_0}$}                            & \multicolumn{2}{c|}{$J_{\gamma_0}$}                            \\ \cline{2-5} 
\multicolumn{1}{|c|}{}                                 & \multicolumn{1}{c|}{Shan.}     & \multicolumn{1}{c|}{min-ent.} & \multicolumn{1}{c|}{Shan.}     & \multicolumn{1}{c|}{min-ent.} \\ \hline
0.000001                                               & \multicolumn{1}{l|}{1.9999784} & 1.9912406                     & \multicolumn{1}{l|}{1.9999784} & 1.9912408                     \\ \hline
0.0001                                                 & \multicolumn{1}{l|}{1.9978744} & 1.9131034                     & \multicolumn{1}{l|}{1.9978744} & 1.9131034                     \\ \hline
0.01                                                   & \multicolumn{1}{l|}{1.7656553} & 1.0459131                     & \multicolumn{1}{l|}{1.7656553} & 1.0459131                     \\ \hline
0.02                                                   & \multicolumn{1}{l|}{1.3794377} & 0.5697659                     & \multicolumn{1}{l|}{1.3794376} & 0.5697659                     \\ \hline
0.0275                                                 & \multicolumn{1}{l|}{0.9593294} & 0.2994331                     & \multicolumn{1}{l|}{0.9593294} & 0.2994330                     \\ \hline
\end{tabular}
	\caption{Comparison between the Shannon entropy and min-entropy lower bounds for certain values of noise, for $I_{\delta}$ and $J_{\gamma}$ protocols. The parameters are chosen to maximize the amount of key certified by these protocols ($\delta = \frac{\pi}{6}$ and $\gamma = 0.0$)}
	\label{tab:comparison}
\end{table}

For these families of Bell inequalities, we observe that both $I_{\delta}$ and $J_{\gamma}$ certify two bits of randomness for Shannon entropy. Additionally, the robustness of the certification for $I_{\delta}$ increases when the value of $\delta$ approaches $\delta = \pi/6$, but the decrease of the value of $\delta$ does not change the maximal certification for that protocol. This behavior has been visualized in Figure~\ref{fig:IDcomparison}. However, in both $I_{\delta}$ and $J_{\gamma}$ families, we were not able to observe a change in the convexity, which was visible in all the previous ones. 

\section{Discussion and Insights} \label{sec:discussion}
The observations of the results above lead to some conclusions about the properties of the Shannon entropy in randomness certification. We were able to observe the general form of Shannon entropy constrained by the NPA optimization results. We were also able to achieve certain gains after applying all possible combinations of the probabilities of results in entropy calculations. Additionally, we have certified randomness for a three-dimensional quantum state.

\subsection{General form for Shannon entropy}
\label{ss:generalform}
The non-linear optimization based on \eqref{eq:nonlinear} yields the value of the Shannon entropy for the provided lower and upper bounds. However, it is also very time-consuming and does not allow for quick certification of Shannon entropy without the need to perform lengthy numerical calculations. However, this value can be analytically lower-bounded.

For $a,b \in \{0,1\}$ and spot-settings $x_0, y_0$, the probability $P(a,b|x,y) \leqslant u_P$, where as $u_P$ we denote an upper bound for the probabilities. As $\sum_{a,b \in \{0,1\}} P(a,b|x,y) = 1$, we can also point out that $u_P \geqslant \frac{1}{n}$, where $n$ is the number of measurement results pairs $(a,b)$, that equals $n=4$ in our analysis (for measurements that allow for more than two possible outcomes, this value can be higher).

One of the features of the Shannon entropy function \eqref{eq:shannon} is that it is Schur-concave\footnote{
A function $f: \mathbb{R}^n \to \mathbb{R}$ is \emph{Schur-concave} if for all $x, y \in \mathbb{R}^n$ such that $x$ is majorized by $y$ (the components of both vectors are sorted in non-increasing order), it holds that $f(x) \ge f(y)$. Equivalently, $f$ is symmetric and satisfies 
\[
(x_i - x_j)\left(\frac{\partial f}{\partial x_i} - \frac{\partial f}{\partial x_j}\right) \le 0
\quad \text{for all } i, j.
\]
Intuitively, a Schur-concave function decreases as its arguments become more unequal, meaning it prefers more balanced distributions of components.}, which implies that it reaches its minimum value for the probability distribution that is the most uneven that the conditions allow. In other words, as few results as possible should appear with the likelihood related to the global probability upper bound. The lowest upper bound in this case will be $u_P = \frac{1}{n}$, where $n$ is the number of different measurement outcomes, but the more biased the results are, the higher $u_P$ will be, and thus entropy will be lower. The distribution that adheres to that allows us to obtain the lower bound for the value of Shannon entropy. Thus, there are going to be exactly $\lfloor 1/u_P \rfloor$ measurement results that will occur with the probability $u_P$, $n-\lfloor 1/u_P \rfloor-1$ results that will not occur ($P(a,b|x,y) = 0$), and one result that will occur with the probability $1 - \lfloor 1/u_P \rfloor u_P$. This way, we can obtain a general form for the lower bound of the Shannon entropy function.

\begin{prop}\label{prop1}
The lower bound for the value of the Shannon entropy function is
\begin{equation}
\begin{split}
	\label{eq:minminchsh}
	H(P) \geqslant - \left\lfloor\dfrac{1}{u_P}\right\rfloor u_P \log_2 \left( u_P \right)\\
	- (1 - \left\lfloor\dfrac{1}{u_P}\right\rfloor u_P) \log_2 \left( 1 - \left\lfloor\dfrac{1}{u_P}\right\rfloor u_P \right),
\end{split}
\end{equation}
where $u_P$ is the upper bound for a single probability in the probability distribution $P$.
\end{prop}

Following \cite{pironio_random_2010}, we employ the relation between the observed CHSH violation value $B$ and the adversary’s optimal guessing probability:
\begin{equation}
	\label{eq:pgminchsh}
    P_{\mathrm{guess}}(a|x) \leqslant \frac{1 + \sqrt{2 - B^2/4}}{2}.
\end{equation}
It has been also shown, that there exists a tight min-entropy bound for local min-entropy $H_{min}(A|X)$, which also constitutes the lower bound for global min-entropy, since $H_{min}(AB|XY) \ge H_{min}(A|X)$ and match: $H_{min}(A|X) \ge 1 - \log_2 \left(1 + \sqrt{2 - B^2/4} \right)$.

\begin{figure}[htbp]
	\includegraphics[width=\linewidth]{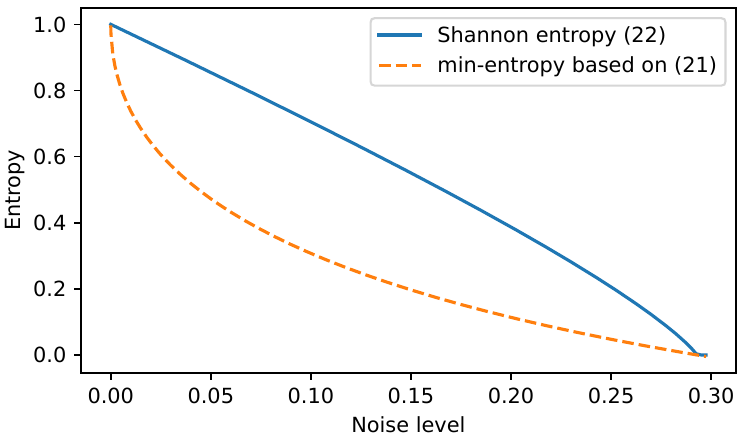}
	\caption{(color online) Shannon and min-entropy lower bounds for CHSH for different levels of noise, based on \eqref{eq:pgminchsh} and \eqref{eq:hminchsh}.}
	\label{fig:hmin}
\end{figure}

Any observed violation above the classical limit guarantees randomness and thus device-independent security. Based on that, we can establish the lower bound for the value of Shannon entropy of a local measurement for CHSH as:
\begin{equation}
\begin{split}
\label{eq:hminchsh}
	H(A|X)^{CHSH} \geqslant - \frac{1 + \sqrt{2 - B^2/4}}{2} \log_2 \left( \frac{1 + \sqrt{2 - B^2/4}}{2} \right) \\ 
	- \left( \frac{1 - \sqrt{2 - B^2/4}}{2} \right) \log_2 \left( \frac{1 - \sqrt{2 - B^2/4}}{2} \right),
\end{split}
\end{equation}
In Figure \ref{fig:hmin} we present how the bounds \eqref{eq:hminchsh} and \eqref{eq:minminchsh} for comparison change depending on the noise level.

\begin{figure}[htbp]
	\includegraphics[width=\linewidth]{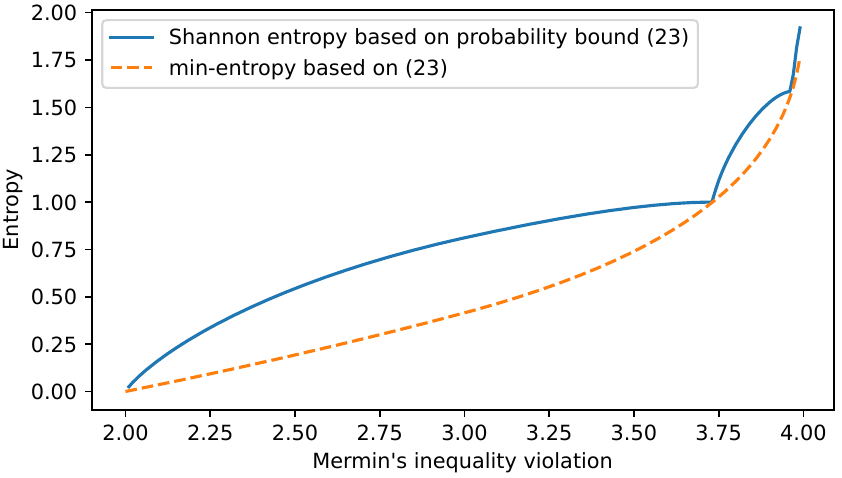}
	\caption{(color online) Shannon and min-entropy lower bounds for CHSH for different Mermin values $2 \leqslant B^{M} \leqslant 4$, based on \eqref{eq:pa1b1_mermin}.}
	\label{fig:hmin2}
\end{figure}

However, as mentioned, the bound based on \eqref{eq:pgminchsh} only describes local measurements of the global state. In \cite{Woodhead_2018}, a tight, analytical guessing probability upper bound for a scenario for three-partite (two honest communication parties and an eavesdropper) Mermin-Bell inequality is presented:
\begin{IEEEeqnarray}{l}
  \label{eq:pa1b1_mermin}
  P_{guess}(ab | e) \IEEEnonumber \\
  \qquad \leqslant \begin{cases}
      \frac{3}{4}  - \frac{B^{M}}{8}
    + \sqrt{3} \sqrt{\frac{B^{M}}{8}
      \left(\frac{1}{2} - \frac{B^{M}}{8}\right)}
    & \text{if } B^{M} \geqslant 3 \\
    \frac{3}{2} - \frac{B^{M}}{4}
    & \text{if } B^{M} \leqslant 3,
  \end{cases} \IEEEeqnarraynumspace
\end{IEEEeqnarray}
where $2 \leqslant B^{M} \leqslant 4$ is the value of the Mermin-Bell inequality achieved in the measurement process. We plot the Shannon entropy lower bound based on \eqref{eq:pa1b1_mermin} and Proposition \ref{prop1}, as well as the min-entropy lower bound based on the same guessing probability upper bound on Figure \ref{fig:hmin2}.

Noticeably, Shannon entropy reaches the same value as min-entropy for two different $B^M$ values, namely for $B^M \approx 3.75$ and $B^M \approx 3.96$. Between $B^M = 2$ and $B^M = 3.75$, our upper bound $u_P > 1/2$, which means that the probability distribution $P = [ u_P, 1-u_P, 0, 0 ]$ and that there are only two possible outcomes, up to $B^M = 3.75$. where $u_P = 1/2$ and Shannon and min-entropy are equal. Between $B^M = 3.75$ and $B^M = 3.94$, there are three possible outcomes and thus, in $B^M = 3.94$, we have three possible outcomes and uniform probability distribution $1/3$ for these outcomes (which again means that the entropy measures are equal). For $B^M > 3.94$ therefore, the number of possible outcomes has to be four. For the Shannon entropy measures obtained using the numerical method from Section~\ref{sec:method} however, there are no common points of min-entropy and Shannon entropy. This is caused by the fact, that we use not just upper, but also lower bounds.

Additionally, we can also make a direct observation considering the probability distributions obtained in the optimization process. We have observed for every analyzed case that a particular probability distribution maximizes the value of Shannon entropy:
\begin{obs}\label{con1}
	The probability distribution that maximizes the value of Shannon entropy and therefore maximizes the randomness for a given certificate is (up to a permutation)
\begin{equation}
    P_H = [l_{-1, -1}, l_{-1, 1}, u_{1, -1}, 1 - l_{-1, -1} - l_{-1, 1} - u_{1, -1}],
\end{equation}
	where $l_{-1, -1}$ is the lower bound for result $a=-1, b=-1$, $l_{-1, 1}$ is the lower bound for result $a=-1, b=1$ and $u_{1, -1}$ is the upper bound for result $a=1, b=-1$.
\end{obs}
This observation has not been proven for every family of Bell inequalities; however, for the analyzed examples in Section~\ref{sec:results}, the observation-based entropy value agrees with the value obtained using nonlinear optimization to 6th precision.

\subsection{Entropy as a composite function}
\label{ssec:entropycomplex}
In Sections~\ref{ssec:chsh}, \ref{ssec:bcn}, and \ref{ssec:i12}, a change in the convexity of the function is noticeable. Shannon entropy is not a function defined on a sequence of intervals; therefore, this behavior may be caused by a different course of the functions that make up the Shannon entropy, especially the course of the NPA-calculated lower and upper probability bounds. To investigate this issue, we have calculated the second derivative of the Shannon entropy (the process is shown in Appendix \ref{appendix:convexity}):
\begin{equation}
	\label{eq:deri2}
\begin{split}
	\frac{\partial^2 H}{\partial p^2} = \frac{\partial H}{\partial l_{-1,-1}} \frac{d^2 l_{-1,-1}}{d p^2} - \frac{1}{\ln 2} \left( \beta_1 \left( \frac{d l_{-1,-1}}{d_{p}} \right)^2 \right.\\
	\left. + \alpha \frac{d l_{-1,-1}}{d p} \left(\frac{d l_{-1,1}}{d p} + \frac{d u_{1,-1}}{d p} \right) \right) + \ldots.
\end{split}
\end{equation}
For these calculations, we were allowed to use the probability distribution from the Observation \ref{con1}, as it is valid for the inequalities described in Section \ref{sec:results}, which include $BC_n$.

\begin{figure}[htbp]
	\includegraphics[width=\linewidth]{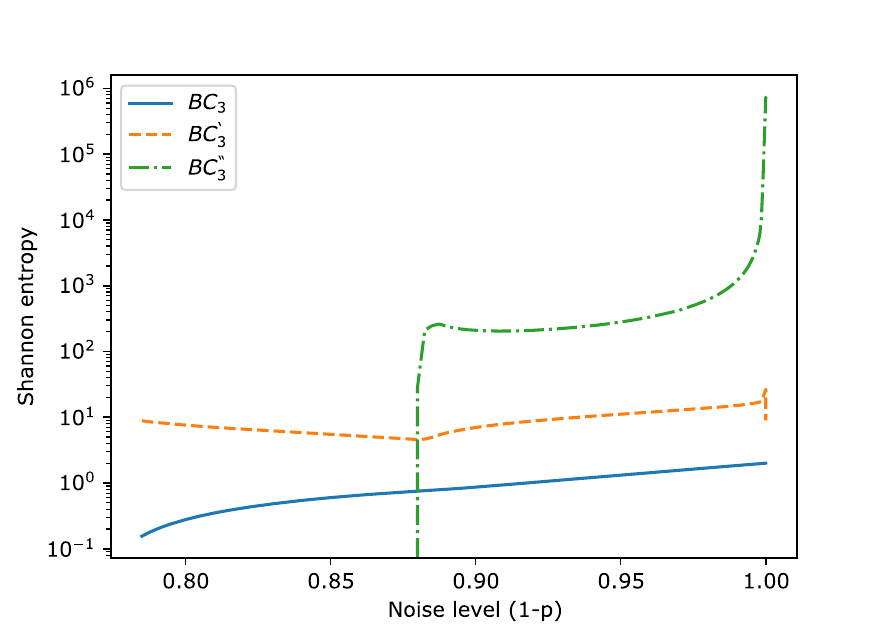}
	\caption{(color online) $BC_3$ function calculated based on Observation \ref{con1} compared with its first and second derivative (in accordance with Appendix \ref{appendix:convexity}).}
	\label{fig:derivative}
\end{figure}

The other coefficients are composed likewise, but with different lower/upper bounds used for derivation. It can be noticed that this derivation is essentially constructed with two addition terms – one positive and one negative. As shown in Figure~\ref{fig:derivative}, on the range of values corresponding to the part with a lower entropy level, the value of the second derivation is lower than 0 (this part of the function is concave), whereas, on the range of values corresponding to the part with higher entropy level, the value of the second derivation is higher than 0 (this part of the function is convex). 

This change in convexity can be essentially explained by the fact that different quantum strategies (in this case, as the measurement settings are fixed -- different measured quantum states) are optimal for different values of the Bell expression (quantum bounds) and different noise levels ($(1-p)B_{max}$). A similar issue has been presented in \cite{Bancal2013MoreRandomness} for optimized guessing probability, where the authors pointed out that mixtures of strategies can achieve higher average guessing probability than any single-strategy curve between those points.

However, this also means that the adversary that is aware of the course of the certified randomness can mix their strategies to exploit the non-convexity to devise a more effective attack. It has also been shown in \cite{Bancal2013MoreRandomness} that if the function of entropy over the Bell inequality violation is not necessarily concave, then for the case with different quantum strategies, the entropy could actually be lower. To ensure that the bound is valid for any mixture of quantum behaviors, a convex hull of the entropy function should be used as the final bound.

Additionally, it can be pointed out that in \eqref{eq:nonlinear}, each value of $l_i$ and $u_i$ is obtained from different NPA optimizations (for a fixed noise level with a Bell inequality used as a certificate), and thus from different quantum behaviors. For computing the Shannon entropy, it would be more convenient to consider a single and fixed quantum behavior. However, the method that minimizes the Shannon entropy over all behaviors does not exist. Let us note, that by considering $l_i$ and $u_i$ from separate optimizations, their constraints can only be relaxed, and thus more pessimistic. What is more, using NPA method itself already introduced another relaxation, as it considers not only all attacks possible in quantum mechanics, but also certain super-quantum attacks. Thus, our method determines the conservative estimate of Shannon entropy value and is a relaxation (as \eqref{eq:npalinear} is a relaxation, the whole method should also be considered a relaxation), where optimization is limited to imposing upper and lower bound constraints, without any further constraints placed on the probability distribution $\{P_i\}$. If we only took into account the allowed quantum behaviors, the entropy would only increase.

\subsection{Additional constraints}
Our NPA calculations were focused on finding the direct lower and upper bounds on the probability of the measurement results alone, as defined for every Bell inequality. This, however, does not exhaust the scope of linear combinations that can be bounded using \eqref{eq:npalinear} and that can be used to further constrain the non-linear \eqref{eq:nonlinear} optimization. The bounded probabilities can thus be combined into 12 other additive linear combinations (as for four different pairs of measurement outcomes, there are exactly $2^4 = 16$ additive linear combinations, and we already used four of them). Two of the combinations, namely the linear combination with no component and the combination with all components, do not provide any additional constraints (as their values are a priori known), and thus we end up with 10 linear combinations that can provide additional, tighter bounds on the sum of the probabilities. These are: $P_{-1,-1} + P_{1,-1}$; $P_{-1,-1} + P_{-1,1}$; $P_{-1,-1} + P_{1,1}$; $P_{-1,1} + P_{1,-1}$; $P_{-1,1} + P_{1,1}$; $P_{1,-1} + P_{1,1}$; $P_{-1,-1} + P_{-1,1} + P_{1,-1}$; $P_{-1,-1} + P_{-1,1} + P_{1,1}$; $P_{-1,-1} + P_{1,-1} + P_{1,1}$; and $P_{-1,1} + P_{1,-1} + P_{1,1}$.
We calculated the additional constraints for modified CHSH \eqref{eq:modchsh} and used these constraints in Shannon entropy non-linear optimization. The lower and upper bounds have been established for subsequent linear expressions. In terms of the mere calculation characteristics, we just needed 100 repetitions using the basin-hopping technique (compared to 1500 repetitions necessary for the previous calculations) to avoid all possible local minima that would disturb the overall result.

\begin{figure}[htbp]
	\includegraphics[width=\linewidth]{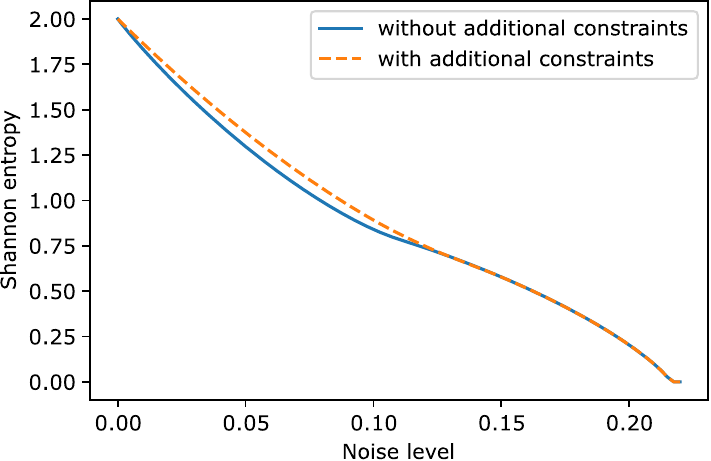}
	\caption{(color online) Lower bounds on Shannon entropy for modified CHSH Bell inequality with and without additional constraints.}
	\label{fig:comparison}
\end{figure}

The results show that with additional constraints, we were able to obtain an advantage to a certain degree. However, the advantage ends for the same noise level, where CHSH starts to have an advantage over the modified CHSH, which is also the moment when the function changes its convexity, as described in Section~\ref{ssec:entropycomplex}.

\subsection{Comparison}
\begin{figure}[htbp]
	\includegraphics[width=\linewidth]{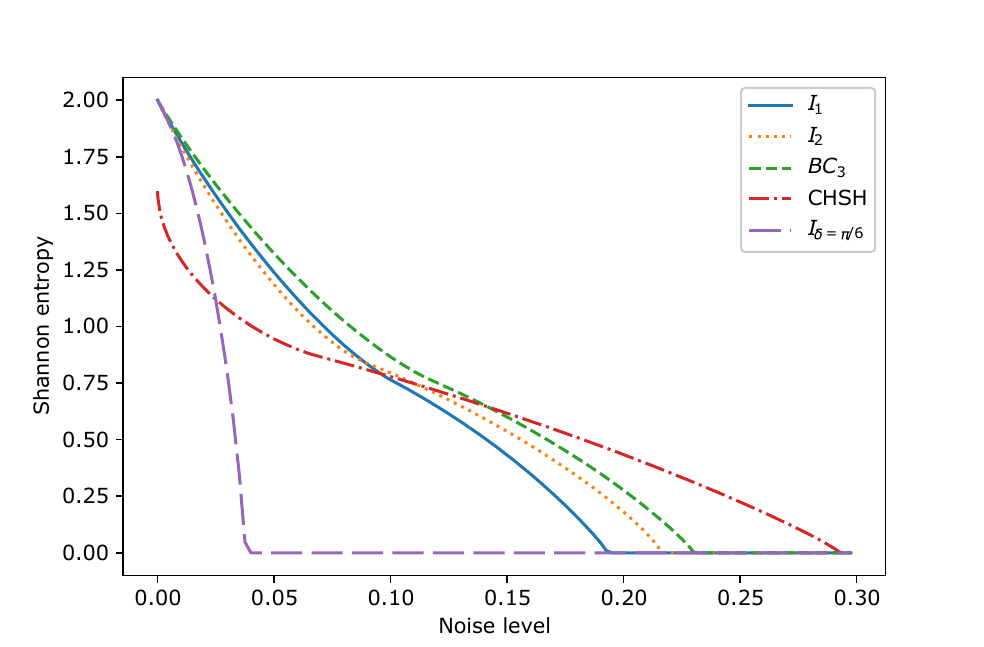}
	\caption{(color online) Lower bounds on Shannon entropy for different Bell inequalities used for randomness certification (in the previous figures), as a function of noise.}
	\label{fig:comp_ost}
\end{figure}

As presented in the previous remarks, the quality of certification differs for certificates applied for different amounts of noise. As presented in Figure~\ref{fig:comp_ost}, there is no candidate for the best certificate in the entire noise range. The robustness of $BC_3$ is dominating up to the noise level of $p = 0.14$ when it is overtaken by the amount of randomness certified by CHSH.

If the user does not want to use $BC_3$, $I_1$ provides the highest randomness up to $p = 0.08$, when $I_2$ starts to have a slight advantage. CHSH beats $I_1$ for $p < 0.095$ and $I_2$ for $p < 0.11$.

\subsection{Three-dimensional scenario}
\label{ssec:3dscenario}
To further investigate the topic of DI randomness certification, we have also pursued a three-dimensional protocol based on the multi-dimensional family of Bell inequalities proposed in \cite{kaniewski_maximal_2019}. This family constitutes a generalisation of the CHSH for $d$-outcome measurements. It also has an interesting property, that the maximal quantum violation is achieved by the maximally entangled states, measured using the projectors that correspond to mutually unbiased bases. Their results are based on the functional due to Buhrman and Massar (BM functional) \cite{PhysRevA.72.052103}. The available number of outcomes $d$ and the number of settings are constrained to odd primes, and the quantum bound can be algebraically calculated using a sum-of-squares decomposition of the Bell expression.

More specifically, we use a Bell operator in the form of \cite{kaniewski_maximal_2019}:
\begin{equation}
	G = \frac{1}{d^3} \sum_{n=1}^{d} \lambda_n \sum_{jk} \omega^{njk} \Pi_n^{0,j} \otimes \Pi_n^{1,k},
\end{equation}
where $\lambda_n$ are complex numbers of unit modulus, $\omega$ are the coefficients for the measurement and $\Pi_n^{0,j}$ and $\Pi_n^{1,k}$ are the observables. The quantum bound we used for the optimization of this inequality is $B = 1.718233$, and the settings are $x = d+1 = 4$ and $y = 2$.

\begin{figure}[!ht]
	\includegraphics[width=\linewidth]{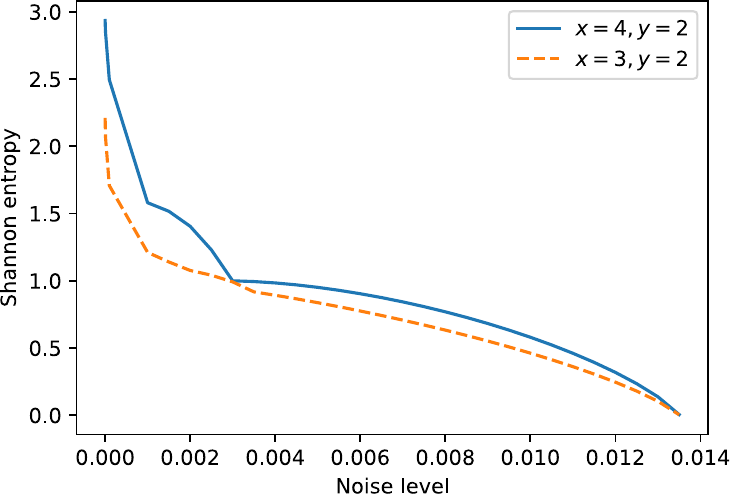}
	\caption{(color online) Shannon certification of randomness using the Bell inequality proposed in \cite{kaniewski_maximal_2019}, for $d = 3$ Werner state, quantum bound $B = 1.718233$, settings $(x=4, y=2)$ and $(x=3, y=2)$ for comparison.}
	\label{fig:threedim}
\end{figure}

We have used our certification method, presented in Section~\ref{sec:exp}. However, as for the three-dimensional quantum state there are 9 possible pair of outcomes, we had to add additional linear-expression constrains to address the other available outcomes, and optimize over 8 independent variables, each one corresponding to the probability of a single pair of values being the outcome of the measurement (with the last probability being the difference between $1$ and the sum of other variables). The outcome of the numerical optimization is presented in Figure~\ref{fig:threedim}.

The results have presented characteristics similar to the two-dimensional Bell certificates that we have investigated -- they presented a change in the concavity of the entropy function -- it changes around the $p = 0.0055$, which is also where the entropy achieves $H^{p = 0.0055} = 1$. For the lowest amount of noise, $p = 1\mathrm{e}{-6}$, which corresponds to the Bell violation value $B^{p = 1\mathrm{e}{-6}} = 1.718231$, almost three bits of randomness are achievable ($H^{p = 1\mathrm{e}{-6}} = 2.93$). However, an important caveat is that the noise tolerance is very low -- entropy converges to zero already around noise level $p = 0.014$ (Bell violation value $B^{p = 0.014} = 1.694178$), which is incomparably worse than for the two-dimensional quantum objects.

\section{Conclusions}
In this paper, we have presented a new randomness certification method based on NPA to establish bounds on the probability distribution and the linear combinations of these probabilities. We have used these bounds as constraints on the non-linear optimization to obtain the value of Shannon entropy and thus conduct the certification. However, our method can be used for any non-convex problem that requires the characterisation of the correlations that arise from locally measuring a single part of a joint quantum system.

Using our method, we compared four families of Bell inequalities, used as a certification for randomness generation: CHSH and modified CHSH, Braunstein-Caves inequalities, $I_1/I_2$, and Wooltorton-Brown-Colbeck inequalities. We have presented which certificate should be used to show the robustness of the QRNG device for different levels of noise. Additionally, we have shown some crucial differences in the certification for Shannon entropy, compared to min-entropy, which is often the subject of such analysis.

In the process, we were able to deduce the form that Shannon entropy takes if it is constrained by NPA-obtained lower and upper probability bounds. We have presented a tight lower bound for Shannon entropy, which allows us to bypass costly non-linear optimization and still estimate a DI security proof. Additionally, in Section \ref{ssec:3dscenario}, we have presented the Shannon entropy certification for a three-dimensional scenario, showing a potential application for randomness extraction for qudits. Overall, the practicality of our results has been underlined by the discussion in Section~\ref{ssec:shannon}, where the significance of the Shannon entropy for randomness estimation has been presented.

\section{Acknowledgments}
We acknowledge the use of a computational server financed by the Foundation for Polish Science (IRAP project, ICTQT, contract no. 2018/MAB/5/AS-1, co-financed by the EU within the Smart Growth Operational Programme). PM was supported by the European Union’s Horizon Europe research and innovation programme under grant agreement No 101080086/NeQST.

We acknowledge the use of a computational server shared by the Immunoinformatics team (Alfaro J., Daghir-Wojtkowiak E., Kallor A., Pałkowski A., Waleron M.) of ICCVS, financed by the EC (Horizon2020 project KATY GA no. 101017453).

NPA optimization was implemented using Python library ncpol2sdpa \cite{Wittek_2015}. We used MOSEK \cite{mosek} as a solver. Non-linear optimization was achieved using SciPy \cite{2020SciPy-NMeth}, using COBYQA as the optimization solver \cite{rago_thesis, cobyqa}.

\onecolumngrid
\begin{appendices}
\section{Second-degree derivation for Shannon entropy}
\label{appendix:convexity}
	For the analysis of the convexity of the Shannon entropy, we had to verify the form of its second-order derivative, concerning functions for lower and upper bounds used by the Conjecture \ref{con1}. As these functions are dependent on the variable $p$, we needed to calculate the derivative of a composite function:
\begin{equation}
	H(l_{-1,-1}(p), l_{-1,1}(p), u_{1,-1}(p)) = H(p),
\end{equation}
where $H$ is the Shannon entropy function, $l_{-1,-1}$ is the lower bound for the probability of receiving a result $(a = -1, b = -1)$, $l_{-1,1}$ is the lower bound for the probability of receiving a result $(a = -1, b = 1)$ and $u_{1,-1}$ is the upper bound for the probability of receiving a result $(a = 1, b = -1)$, as in the Conjecture \ref{con1}.

Based on the chain rule for differentials:
\begin{equation}
	\left.\frac{\partial H}{\partial p}\right|_{l_{-1,-1}, l_{-1,1}, u_{1,-1}} = \left.\frac{\partial H}{\partial l_{-1,-1}}\right|_{l_{-1,1}, u_{1,-1}} \frac{d l_{-1,-1}}{d p} + \left.\frac{\partial H}{\partial l_{-1,1}}\right|_{l_{-1,-1}, u_{1,-1}} \frac{d l_{-1,1}}{d p} + \left.\frac{\partial H}{\partial u_{1,-1}}\right|_{l_{-1,-1}, l_{-1,1}} \frac{d u_{1,-1}}{d p}.
\end{equation}

The values of $\frac{d l_{-1,-1}}{d p}$, $\frac{d l_{-1,1}}{d p}$ and $\frac{d u_{1,-1}}{d p}$ were obtained using central differences method, as these have to be based on the results of the NPA optimization. The Shannon entropy formula is based on Observation \ref{con1}:
\begin{equation}
\begin{aligned}
	H = - l_{-1,-1} \log_2 (l_{-1,-1}) - l_{-1,1} \log_2 (l_{-1,1}) - u_{1,-1} \log_2 (u_{1,-1})\\
	- (1-l_{-1,-1}-l_{-1,1}-u_{1,-1}) \log_2(1-l_{-1,-1}-l_{-1,1}-u_{1,-1}).
\end{aligned}
\end{equation}
Thus, partial derivatives with respect to lower and upper bounds equal:
\begin{equation}
\begin{aligned}
\left.\frac{\partial H}{\partial l_{-1,-1}}\right|_{l_{-1,1}, u_{1,-1}} = \log_2 \left(\frac{1 - l_{-1,-1} - l_{-1,1} - u_{1,-1}}{l_{-1,-1}}\right),\\
\left.\frac{\partial H}{\partial l_{-1,1}}\right|_{l_{-1,-1}, u_{1,-1}} = \log_2 \left(\frac{1 - l_{-1,-1} - l_{-1,1} - u_{1,-1}}{l_{-1,1}}\right),\\
\left.\frac{\partial H}{\partial u_{1,-1}}\right|_{l_{-1,-1}, l_{-1,1}} = \log_2 \left(\frac{1 - l_{-1,-1} - l_{-1,1} - u_{1,-1}}{u_{1,-1}}\right).
\end{aligned}
\end{equation}

The second derivative is now simply a sum of derivatives of the components of the first derivative. In other words:
\begin{equation}
\label{eq:secder}
	\frac{\partial^2 H}{\partial p^2} = \frac{\partial}{\partial p} \left(\frac{\partial H}{\partial l_{-1,-1}} \frac{d l_{-1,-1}}{d p}\right) +  \frac{\partial}{\partial p} \left(\frac{\partial H}{\partial l_{-1,1}} \frac{d l_{-1,1}}{d p}\right) +  \frac{\partial}{\partial p} \left(\frac{\partial H}{\partial u_{1,-1}} \frac{d u_{1,-1}}{d p}\right).
\end{equation}

A single component equals:
\begin{equation}
\begin{aligned}
	\frac{\partial}{\partial p} (\frac{\partial H}{\partial l_{-1,-1}} \frac{d l_{-1,-1}}{d p}) = \frac{d^2 l_{-1,-1}}{d p^2} \frac{\partial H}{\partial l_{-1,-1}} + \left(\frac{d l_{-1,-1}}{d p}\right)^2 \frac{\partial^2 H}{\partial l_{-1,-1}^2} \\
	+ \frac{d l_{-1,-1}}{d p} \left( \frac{d l_{-1,1}}{d p} \frac{\partial^2 H}{\partial l_{-1,1} \partial l_{-1,-1}} + \frac{d u_{1,-1}}{d p} \frac{\partial^2 H}{\partial u_{1,-1} \partial l_{-1,-1}} \right).
\end{aligned}
\end{equation}
Once again, $\frac{d l_{-1,-1}}{d p}$, $\frac{d l_{-1,1}}{d p}$ and $\frac{d u_{1,-1}}{d p}$ (as well as $\frac{d^2 l_{-1,-1}}{d p^2}$) were obtained using central differences. The other subcomponents equal:

\begin{equation}
\begin{aligned}
	\frac{\partial^2 H}{\partial l_{-1,-1}^2} = - \frac{1}{\ln 2} \left(\frac{1 - l_{-1,1} - u_{1,-1}}{1 - l_{-1,-1}^2 - l_{1,-1} - u_{1,-1}}\right),\\
	\frac{\partial^2 H}{\partial l_{-1,1} \partial l_{-1,-1}} = \frac{\partial^2 H}{\partial u_{1,-1} \partial l_{-1,-1}} = - \frac{1}{\ln 2} \left(\frac{1}{1 - l_{-1,-1} - l_{1,-1} - u_{1,-1}}\right).
\end{aligned}
\end{equation}
Additionally, we denote:
\begin{equation}
\begin{aligned}
	\alpha = \frac{1}{1 - l_{-1,-1} - l_{-1,1} - u_{1,-1}},\\
	\beta_1 = \frac{1 - l_{-1,1} - u_{1,-1}}{1 - l_{-1,-1}^2 - l_{-1,1} - u_{1,-1}},\\
	\beta_2 = \frac{1 - l_{-1,-1} - u_{1,-1}}{1 - l_{-1,-1} - l_{-1,1}^2 - u_{1,-1}},\\
	\beta_3 = \frac{1 - l_{-1,-1} - l_{-1,1}}{1 - l_{-1,-1} - l_{-1,1} - u_{1,-1}^2}.
\end{aligned}
\end{equation}
The previous subcomponents are thus reduced to
\begin{equation}
\begin{aligned}
	\frac{\partial^2 H}{\partial l_{-1,-1}^2} = - \frac{\beta_1}{\ln 2}, \hspace{3em}
	\frac{\partial^2 H}{\partial l_{-1,1} \partial l_{-1,-1}} = \frac{\partial^2 S}{\partial u_{1,-1} \partial l_{-1,-1}} = - \frac{\alpha}{\ln 2}.
\end{aligned}
\end{equation}
The full component can therefore be reduced to
\begin{equation}
\begin{aligned}
	\frac{\partial}{\partial p} (\frac{\partial H}{\partial l_{-1,-1}} \frac{d l_{-1,-1}}{d p}) = \frac{d^2 l_{-1,-1}}{d p^2} \log_2 \frac{1}{\alpha l_{-1,-1}} - \frac{1}{\ln 2} \left( \beta_1 \left( \frac{d l_{-1,-1}}{d p}\right)^2 + \alpha \frac{d l_{-1,-1}}{d p} \left( \frac{d l_{-1,1}}{d p} + \frac{d u_{1,-1}}{d p} \right) \right).
\end{aligned}
\end{equation}
We transform the remaining components of \eqref{eq:secder}, which leads to \eqref{eq:deri2}.

\end{appendices}
\twocolumngrid

\bibliographystyle{plainnat}
\bibliography{shannon}

\end{document}